\newif\ifjcp
\jcptrue
\ifjcp
    \documentclass[twocolumn,aip,jcp,reprint,floatfix]{revtex4-1}
\else
    \documentclass[journal=jacsat,manuscript=article]{achemso}
    \setkeys{acs}{articletitle = true}
    \newcommand{\onlinecite}[1]{\hspace{-1 ex} \nocite{#1}\citenum{#1}} 
\fi

\usepackage[T1]{fontenc}
\usepackage{natbib}
\usepackage{graphicx}
\usepackage{float}
\usepackage{physics}
\usepackage{amsmath}
\usepackage{xcolor}
\usepackage{capt-of}

\newcommand{\toadd}[1]{{#1}}
\newcommand{\toremove}[1]{{}}

\newcommand{\setinfo}{%
    \title{Scalable and predictive spectra of correlated molecules with moment truncated iterated perturbation theory}
    \author{Oliver J. Backhouse}%
    \affiliation{Department of Physics, King's College London, Strand, London WC2R 2LS, U.K.}%
    \author{Alejandro Santana-Bonilla}%
    \affiliation{Department of Physics, King's College London, Strand, London WC2R 2LS, U.K.}%
    \author{George H. Booth}%
    \email{george.booth@kcl.ac.uk}%
    \affiliation{Department of Physics, King's College London, Strand, London WC2R 2LS, U.K.}%
}

\ifjcp
    \renewenvironment{figure}{%
        \begin{figure*}%
    }{%
        \end{figure*}%
        \ignorespacesafterend%
    }
\else
\fi

\ifjcp\else
    \setinfo
\fi

\begin{document}

\ifjcp
    \setinfo
\fi

% MANUSCRIPT GUIDELINES:
% Letters are short articles that report results whose immediate availability to the scientific community is deemed important. 
% The preferred maximum length for each Letter is 2500 words or the equivalent (8– 10 double-spaced typewritten pages of text, 3–4 figures, and 1–2 schemes/illustrations). 
% A brief abstract of no more than 150 words should be included; instructions for preparing an appropriate abstract may be found below. 
% Special effort will be made to expedite the reviewing and the publication of JPC Letters submissions. 
% Thus, authors should ensure that manuscripts are in final, error-free form when submitted. 
% Letters must contain a Table of Contents (TOC)/Abstract graphic as part of the manuscript.

\begin{abstract}
    A reliable and efficient computation of the entire single-particle spectrum of correlated molecules is an outstanding challenge in the field of quantum chemistry, with standard density functional theory approaches often giving an inadequate description of excitation energies and gaps. In this work, we expand upon a recently-introduced approach which relies on a fully self-consistent many-body perturbation theory, coupled to a non-perturbative truncation of the effective dynamics at each step. We show that this yields a low-scaling and accurate method across a diverse benchmark test set, capable of treating moderate levels of strong correlation effects, and detail an efficient implementation for applications up to $\sim1000$ orbitals on parallel resources. We then use this method to characterise the spectral properties of the artemisinin anti-malarial drug molecule, resolving discrepancies in previous works concerning the active sites of the lowest energy fundamental excitations of the system.
\end{abstract}

\ifjcp
    \maketitle
\fi

%
% INTRODUCTION
%

The single-particle Green's function is a compact object, but with a vast wealth of information about an interacting quantum system.
Insight into the both the ground state static quantities, as well as the excitation spectrum is accessible, which is key in governing the response, transport and optical properties of the system, as well as being directly accessed through experimental forward and inverse photoelectron spectroscopy.
It is unsurprising therefore that since early in the development of electronic structure theory, this single-particle Green's function has often been cast as a central quantum variable.
These include perturbative approximations such as the outer-valence Green's function (OVGF)\cite{VonNiessen1984}, algebraic diagrammatic construction (ADC)\cite{Schirmer1983,Trofimov1995,Trofimov1999} and the $GW$ approximation\cite{Hedin1960}, as well as non-perturbative methods such as dynamical mean-field theory (DMFT)\cite{Metzner1989,Zhang1993,Georges1996,Sun2016}.
Furthermore, methods built around equation-of-motion formalisms for excitations are also closely related to Green's function formalisms, allowing for connections to be developed between existing theories\cite{Stanton1994,Krylov2008,Lange2018}.

In this letter we demonstrate and expand a new approach within this framework which has been recently introduced, auxiliary second-order Green's function (AGF2)\cite{Backhouse2020a,Backhouse2020b}.
This method exhibits a number of appealing features, as well as also building connections between established methods, to bring a new perspective on these approaches.
This includes a favourable $\mathcal{O}[N^5]$ computational scaling with efficient large-scale parallelism, an almost entirely reference-independent and fully self-consistent algorithm, and access to the entire energy-dependence of the excitation spectrum, rather than a state-specific approach to excitations.
Previous work has shown AGF2 to be an accurate and competitive method for both static energetics, as well as charged excitations.
Here, we detail an efficient algorithm for its evaluation which is implemented in a publicly available codebase, allowing access to significant system sizes over large test sets, as well demonstrating its efficient application to open problems in theoretical molecular spectroscopy.

The AGF2 method was originally proposed as a reformulation of its `parent' method, iterated second-order Green's function (GF2)\cite{Holleboom1990,VanNeck2001,Dahlen2005,Phillips2014}.
In this reformulation, it avoided the numerically cumbersome description of an explicit continuous dynamical variable, by describing its effect as static auxiliary degrees of freedom.
This necessitated the algorithmic addition of a renormalisation group-inspired compression of the effective dynamics of the self-energy at each step, which nevertheless conserved key properties of the resulting self-energy via its spectral moments.
It was subsequently realised that in addition to this compression ensuring a favorable numerical cost, for certain well-defined truncations it in fact also substantially improved the spectral accuracy of the resulting theory across a range of systems.
This observation is in keeping with the common feature in perturbative Green's functions methods, that full self-consistency of the propagators generally has a deleterious effect on spectral properties, and thus a well-balanced self-consistency in AGF2 can be motivated by the efficient truncation of these dynamics, despite the loss of a rigorously conserving approximation\cite{Welden2015,VanSetten2013,VanSetten2015,VanSchilfgaarde2005,Kaplan2016}.

However, the AGF2 method is increasingly also being seen as a self-consistent extension of the ADC(2) method, a popular and low-scaling approach to determine accurate excitations in quantum chemistry.
This approach builds an effective Hamiltonian from an equation-of-motion formalism around the M{\o}ller-Plesset second-order perturbative ground state, including all (bare) second-order diagrams\cite{Schirmer1983,Trofimov1995,Trofimov1999,Dempwolff2019,Dempwolff2020,Banerjee2019,Banerjee2021,Herbst2020}.
This effective Hamiltonian can be diagonalised for extremal eigenvalues.
The AGF2 method can be considered as a way to compress this effective Hamiltonian of ADC at each step, and use the result to renormalise or `dress' the propagators in the perturbative expansion.
This self-consistency results in a resummation of diagrams to all orders and an insensitivity to the choice of reference state.
Meanwhile, the systematic compression allows for the entire effective Hamiltonian to be diagonalised, avoiding the need to rigorously separate particle and hole excitations, and ensuring simultaneous access to excitations over all energy scales.

Previous work has referred to this efficient compression (or renormalisation) of the auxiliary space representing the effective self-energy as AGF2(1,0), to distinguish it from other such truncations.
This ensures that at each iteration, the first-order particle and hole spectral moments of the full second-order self-energy are conserved in the truncation.
We will exclusively consider this truncation in this work (and therefore just refer to it as AGF2), and demonstrate that this reduces to a particularly efficient algorithm, with favorable parallelism which can allow a comfortable treatment of 1000 orbitals.
This is implemented within the {\tt PySCF} open-source simulation package\cite{pyscf,pyscf2}.
This resulting scalable implementation allows for benchmarking on the widely used `$GW100$' molecular test set for charged excitations in large basis sets \cite{VanSetten2015}, compared to the established and similarly scaling ADC(2)\cite{Schirmer1983,Trofimov1995,Trofimov1999} and $GW$ methods\cite{Hedin1960,Zhu2021}, as well as higher-scaling equation-of-motion coupled-cluster methods\cite{Monkhorst1977,Stanton1993,Stanton1994,Lange2018}.
Finally, we apply the AGF2 approach to an open problem of the excitation levels in the artemisinin drug molecule, where light is shed on the conflicting results over the spatial location of the assignment of peaks in previous studies in the literature.

We briefly review the AGF2 method to give context to the efficient algorithm presented later and to clarify emerging connections to the ADC method, with a more detailed background available in Refs.~\onlinecite{Backhouse2020a,Backhouse2020b}.
The propagator in the frequency domain, corresponding to the removal or attachment of an electron from a mean-field reference, is given by the single-particle Green's function,
\begin{align} \label{eq:gf_hf}
    G_0(\omega) = \big[ \omega I - F \big]^{-1},
\end{align}
where $F$ is the generalised Fock matrix, with elements
\begin{align} \label{eq:fock_matrix}
    F_{pq} = h_{pq} + \sum_{rs} \big[ (pq|rs) - \frac{1}{2} (ps|rq) \big] D_{rs},
\end{align}
where $h$ is the one-electron (kinetic and electron-nuclear) Hamiltonian, $D$ the one-body reduced density matrix, and $(pq|rs)$ the electronic repulsion integrals.
Poles in the Green's function correspond to the excitation energies to electron-attached/removed states. % at energies $\omega=\epsilon$.
The one-particle Green's function can be dressed from this mean-field description, due to the presence of explicit electron interactions and resulting correlation effects, according to the Dyson equation
\begin{align} \label{eq:dyson_eq}
    G(\omega) = G_0(\omega) + G_0(\omega) \Sigma(\omega) G(\omega),
\end{align}
where $\Sigma(\omega)$ is the frequency-dependent self-energy.
In AGF2, we take this self-energy to be built up iteratively from all bare second-order diagrams (direct and exchange), given by
\begin{align} \label{eq:mp2_self_energy}
    \Sigma^{\mathrm{(2)}}_{pq}(\omega) &= \sum_{ij}^{\mathrm{occ}} \sum_{a}^\mathrm{vir} \frac{(pi|ja) [ 2 (qi|ja) - (qj|ia) ]}{\omega - \epsilon_i - \epsilon_j + \epsilon_a}
    \\
    \nonumber
                        &+ \sum_{ab}^{\mathrm{vir}} \sum_{i}^\mathrm{occ} \frac{(pa|bi) [ 2 (qa|bi) - (qb|ai) ]}{\omega - \epsilon_a - \epsilon_b + \epsilon_i},
\end{align}
where we can consider the first term spanning the $2h1p$ space to be the `lesser' self-energy ($\Sigma_{pq}^{(2),<}$), and the second, spanning the $1h2p$ space to be the `greater' self-energy ($\Sigma_{pq}^{(2),>}$).
\toadd{
    We can exactly expand the effect of this dynamical self-energy as a set of static auxiliary degrees of freedom. This can be considered an inversion of the `downfolding' procedure via L{\"o}wdin partitioning, where degrees of freedom can be integrated out and exactly represented as a dynamical potential \cite{Lowdin1962,Loos2020b,Hirata2015,Hirata2017}.
To see this, we note that a fully causal self-energy} such as the one in Eq.~\ref{eq:mp2_self_energy} can be rewritten in the form
\begin{align} \label{eq:self_energy}
    \Sigma_{pq}(\omega) = \sum_{k} \frac{v_{pk} v_{qk}^\dagger}{\omega - E_k}.
\end{align}
\toadd{This allows one to describe a renormalization of the propagator due this dynamical self-energy as an eigenvalue problem, as a static formulation of Eq.~\ref{eq:dyson_eq}, as}
\begin{align} \label{eq:dyson_eq_eig}
    \begin{bmatrix} & F & v & \\ & v^\dagger & \mathrm{diag}(E) & \end{bmatrix} \phi = \lambda \phi.
\end{align}
In our case, the first sector of the matrix corresponds to $1h$ (occupied/hole) and $1p$ (virtual/particle) states (indexed by the general physical space orbitals $p,q$), and the second to $2h1p$ and $1h2p$ states (given by compound indices $ija$ and $iab$).
The eigenvectors $\phi$ and eigenvalues $\lambda$ give the exact pole structure of the propagator having been dressed by the second-order diagrams corresponding to Eq.~\ref{eq:mp2_self_energy}. \toadd{Using this dressed propagator, we can then form a new self-energy, updating the poles and residues of Eq.~\ref{eq:mp2_self_energy}, to allow the procedure to be in principle iterated to convergence, and thus resumming the effect of the diagrams to all orders. This would result in a frequency-free, but physically identical reformulation of the GF2 method \cite{VanNeck2001}. However, the dimensionality of the eigenvalue problem would grow exponentially with iterations, necessitating a compression of this auxiliary space. The specifics of this compression can materially change results, and previous work has shown that our physically-motivated compression based on constraints from the spectra moments of Eq.~\ref{eq:mp2_self_energy} can have a dramatic improvement on the resulting overall spectral properties of the method compared to the `parent' GF2 method\cite{Backhouse2020b}. We detail an efficient formulation of this compression later.}

\toadd{We can also relate the approach mathematically to the popular ADC(2) approach.} This (first iteration) AGF2 matrix has an identical form to the `Dyson' ADC(2) matrix, which similarly includes these second-order diagrams, built on a Hartree--Fock reference \footnote{In `Dyson' ADC(2), there would additionally be a perturbative relaxation of the one-body density matrix. However, in AGF2, this density matrix is self-consistently relaxed through the iterations \cite{Schirmer1998}}.
The dimension of this matrix scales as $\mathcal{O}[N^3]$, where $N$ is the number of orbitals in the system. To render this tractable, ADC(2) therefore separates the $2h1p$ and $1h2p$ contributions to the self-energy (the first and second terms in Eq.~\ref{eq:mp2_self_energy}, respectively), in order to ensure that the extremal eigenvalues to these separated eigenvalue problems correspond to the lowest energy excitations (IPs and EAs), which can be efficiently solved via iterative eigensolvers (e.g. Davidson algorithm) at $\mathcal{O}[N^5]$ cost.
To account for this separation, taking IP-ADC(2) as an example, the $1h$ space gains an additional term consistent through second-order perturbation theory describing this neglected mixing between the particle and hole contributions to the self-energy\cite{Schirmer1983,Schirmer1998,Banerjee2019,Mester2018}.
This is sometimes termed a `non-Dyson' ADC approach, and is now standard in the community, with all ADC results in this work corresponding to this non-Dyson ADC approximation where particle and hole sectors are separated.

\begin{figure}[bt!]
    \centering
    \ifjcp
        \includegraphics[width=1.16\columnwidth]{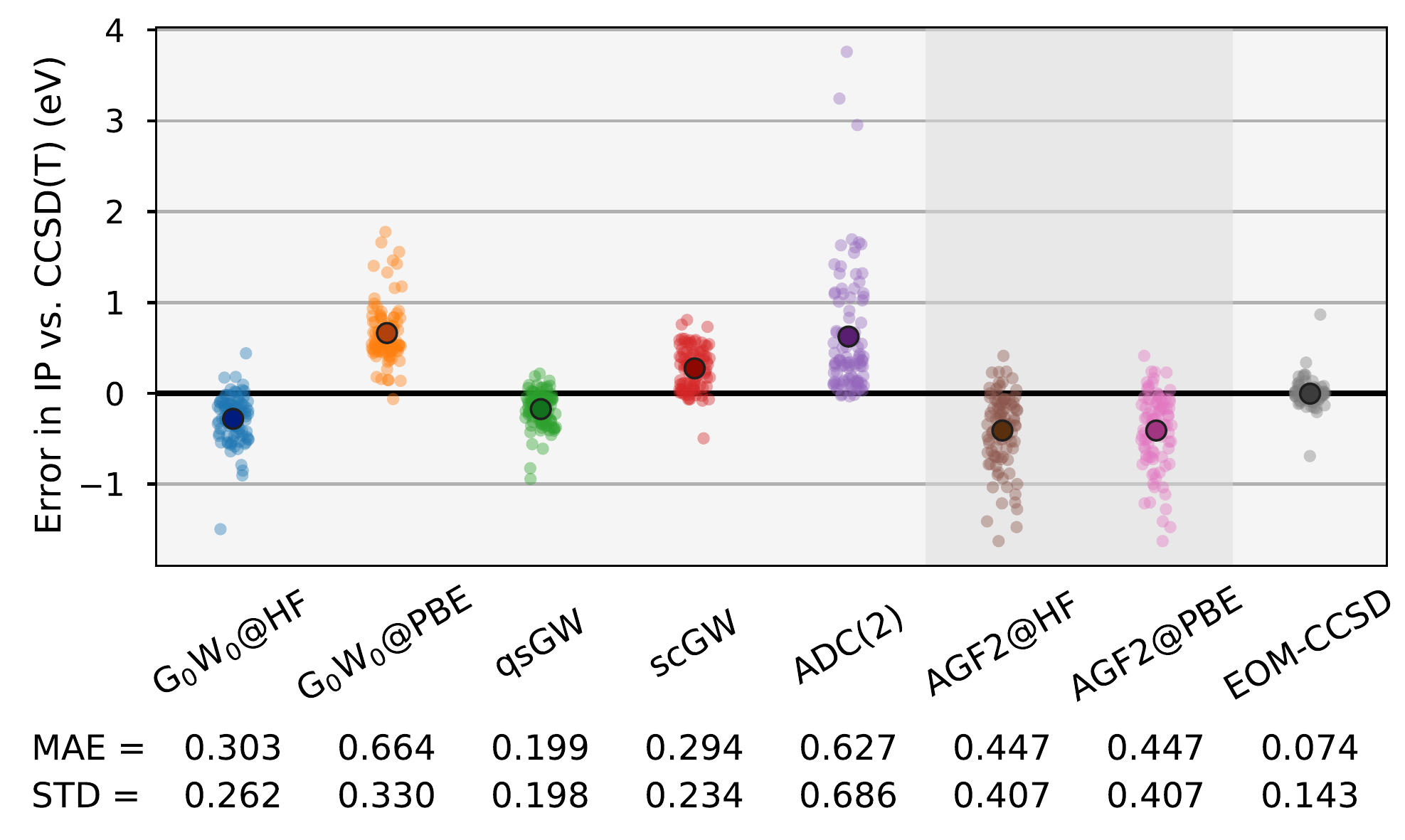}
        \includegraphics[width=0.84\columnwidth]{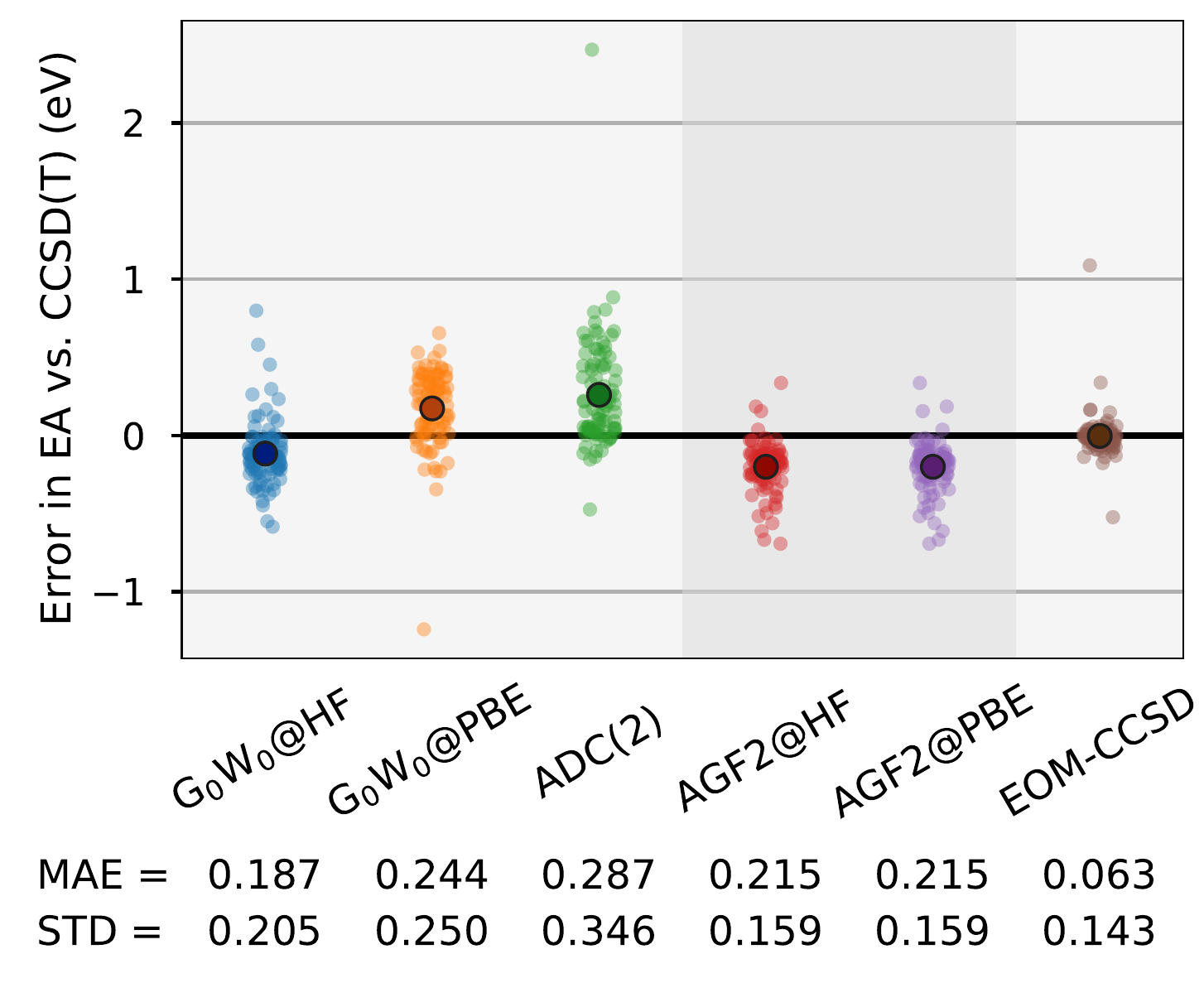}
    \else
        \includegraphics[width=0.57\textwidth]{gw100_ip_ext.pdf}
        \includegraphics[width=0.41\textwidth]{gw100_ea.pdf}
    \fi
    \caption{
    Errors in the computed IP (left) and EA (right) for AGF2 and a series of other established methods, across the GW100 benchmark set in a def2-TZVPP basis.
    %Molecules containing the atoms Rb, Ag, I, Cs, Au or Xe, and additionally hexafluorobenzene, have been omitted.
    Under each method is the mean absolute error (MAE) and standard deviation of the signed errors (STD), while the solid point in each methods scatter indicates the mean (signed) error across the values. Errors are calculated by comparison to benchmark CCSD(T) values, for which separate $N+1$, $N$, and $N-1$ electron systems were computed.
    }
    \label{fig:gw100}
\end{figure}

%requires the use of iterative eigensolvers to target interior eigenvalues (IPs and EAs), and so it necessitates the separation of the $2h1p$ and $1h2p$ self-energies (the first and second terms in Eq.~\ref{eq:mp2_self_energy}, respectively)\cite{TODO}.
%Taking IP-ADC(2) as an example, to account for this, the $1h$ space gains an additional term consistent through second-order perturbation theory, which is an approximation of the effect of the neglected $1h2p$ self-energy\cite{TODO}. 
%This is termed a `non-Dyson approach', and allows the computation of IPs and EAs with e.g. the Davidson algorithm, as a full diagonalization of the matrix to obtain the entire eigenspectrum of $\phi,\lambda$ would scale as $N^9$ with system size.
In AGF2 we instead employ a compression of the full frequency-dependence of $\Sigma^{(2)}(\omega)$, in order to permit the full diagonalisation of the resulting effective second-order matrix of Eq.~\ref{eq:dyson_eq_eig}.
This is achieved via a rotation and subsequent truncation of the $2h1p$/$1h2p$ sector of the matrix, according to a systematic criteria to preserve the spectral moments of the resulting self-energy or propagators\cite{Weikert1996}.
Rather than only obtaining extremal parts of the spectrum, we therefore obtain a coarse-grained representation of the entire spectrum, whilst avoiding the requirement to separate the hole and particle self-energy contributions as performed in ADC.
Furthermore, this approach permits a self-consistency via Eq.~\ref{eq:mp2_self_energy} in which $i,j,a,b$-labelled states become the renormalised orbitals $\phi$, termed `quasi-molecular orbitals' (QMOs), instead of the original mean-field molecular orbitals (MOs).
This iteratively includes diagrammatic contributions at higher orders, summing these diagrams (in the compressed representation) to infinite order.
\toadd{
    As such, the propagator is iteratively renormalised with all possible insertions of the irreducible second-order diagrams, with the compression representing a coarse-graining of the full resulting self-energy over their time orderings.
}
An additional relaxation of the one-body density matrix due to the correlations is included each iteration, via these correlation driven modification to the effective (Dyson) orbitals, along with a chemical potential $\mu$ which must be maintained throughout in order to ensure that there is the correct number of electrons in the resulting non-idempotent physical-space density matrix.
We note that similar self-consistent procedures exist in the literature, including the real-frequency GF2 approach of Piers et. al\cite{Piers2002}, as well as recent related work in the nuclear physics community on a truncated self-consistent ADC\cite{Cipollone2013,Soma2014,Barbieri2017,Raimondi2018,Barbieri2018,Porro2021}.

Before describing the efficient $\mathcal{O}[N^5]$-scaling implementation of the chosen compression algorithm, we demonstrate the efficacy of this truncated self-consistent AGF2 approach on the $GW100$ test set established for the benchmarking of charged excitations\cite{VanSetten2015}.
In Fig.~\ref{fig:gw100}, we present the $GW100$ test set errors for IPs and EAs of AGF2, correlating all electrons in a def2-TZVPP basis set, compared to the similarly-scaling ADC(2), the $\mathcal{O}[N^6]$-scaling EOM-CCSD\cite{Monkhorst1977,Stanton1993,Stanton1994} and \toadd{various $GW$-derived methods, which scale as $\mathcal{O}[N^4]$ in the most common implementations \cite{Hedin1960,Hybertsen1985,Aryasetiawan1998,VanSetten2013,Kaplan2016,Bintrim2021}, but lower-scaling $\mathcal{O}[N^3]$ state-of-the-art implementations are emerging \cite{Kresse2016,Duchemin2021}.}
For the AGF2 and $G_{0}W_{0}$ methods, we present data starting from both Hartree--Fock (HF) and PBE references, as the $G_{0}W_{0}$ approach exhibits a significant dependence on this initial reference, and PBE is a commonly used choice\cite{Caruso2012}. The $G_{0}W_{0}$ calculations were performed on the {\tt Fiesta} code\cite{Blase2011a,Blase2011b,Duchemin2020}. Since effective core potentials (ECPs) are not currently available in this codebase, molecules in the set which contained the atoms Rb, Ag, I, Cs, Au or Xe were removed.
%The molecules in the set which contained the atoms Rb, Ag, I, Cs, Au or Xe were removed, as effective core potentials (ECPs) are not currently available in the {\tt Fiesta} code used for the $G_{0}W_{0}$ calculations\cite{Blase2011a,Blase2011b}, and 
Furthermore, convergence was unsuccessful with default AGF2 options for the hexafluorobenzene molecule, and so this molecule was also removed from the test set.
\toadd{In order to alleviate the reference dependence of $G_{0}W_{0}$, we also consider quasiparticle self-consistent (qs) and fully self-consistent (sc) $GW$, where the propagators are renormalized with a static approximation or the fully dynamical self-energy respectively \cite{VanSchilfgaarde2005,Holm1998}. These results are available for the IP only from the $GW100$ database \cite{VanSetten2015,Caruso2016}.}
The AGF2 and ADC(2) calculations were performed using the {\tt PySCF} simulation package\cite{pyscf,pyscf2,Backhouse2020a,Backhouse2020b,Banerjee2021}.
The EOM-CCSD values were taken from Ref.~\onlinecite{Lange2018}.
The reference CCSD(T) values were calculated using the {\tt ORCA} program package\cite{Neese2012,Neese2020}.
%\toadd{
%    For the IP, we also include data at the quasiparticle self-consistent (qs) and fully self-consistent (sc) $GW$ levels of theory, which were obtained from the $GW100$ database\cite{VanSetten2015}.
%    Values for the EA at these levels of theory are not available in the database.
%}

Figure~\ref{fig:gw100} shows the distribution of errors in the IPs and EAs, along with mean absolute errors (MAE) and standard deviations of the signed error (STD) for the remaining 94 molecules, in comparison to $\Delta$-CCSD(T) benchmark values.
In keeping with conclusions from previous work over smaller test sets, we see that AGF2 improves on the accuracy of these low-energy excitations compared to the similarly scaling ADC(2).
The quality of the ADC(2) results is particularly impacted by the existence of large outliers.
In the case of IPs their identities are Cu$_2$, CuCN and MgO, whilst for the EA the single large outlier is TiF$_4$.
These systems are examples of stronger correlation, where the MP2 ground-state of ADC(2) is inadequate, and showing the necessity for higher order diagrammatic contributions, either through a higher order level of theory or resummation of low-order perturbation theory as performed exactly in coupled-cluster, or approximately in AGF2.
Van Setten \textit{et al.} note that the three problematic cases for the IP belong to a subset of the $GW100$ set which present significant quasiparticle renormalization, characterized by large values for the derivative $\frac{\partial\Sigma(\omega=\epsilon)}{\partial\omega}$.
This \toadd{suggests} the presence of stronger correlation effects, \toadd{and in $G_0W_0$ was also found to result in non-uniqueness and convergence difficulties in the solution to the quasiparticle equation\cite{VanSetten2015}.}
Further analysis shows that the character of the IPs for these systems agrees between the self-consistent methods EOM-CCSD and AGF2, while deviating qualitatively in their assignment with ADC(2) and ADC(3).
For CuCN and MgO, these IPs are not HOMO-like in their character, and have a relatively low quasiparticle weight, which likely explains the difficulty in capturing this physics in non-self-consistent approaches.
\toadd{
    Due in part to the lack of an iterative eigensolver or technical convergence parameters to consider in AGF2, we find that the method is robust and relatively insensitive to technical details compared to other `fully-dynamical' Green's function approaches, which generally require choices of grids in various domains and can observe multiple solutions \cite{Pokhilko2021,Loos2018,Pina2021}.
    The present AGF2 calculations were converged using the default settings in {\tt PySCF}, and no manual selection of states or physical solutions is required.
    Whilst AGF2 makes use of DIIS and damping to accelerate and stabilise the self-consistency \cite{Pulay1980}, beyond the basis set there are essentially no such convergence parameters which change the resulting solutions.
}

\toadd{Finally, the use of a self-consistent framework for AGF2 results in reference-independent values for the IP and EAs, in contrast to the widely used $G_{0}W_{0}$ method for charged excitations, which retains a dependence on the initial choice of exchange-correlation functional in the reference state.
This is found to manifest in considerable differences in the performance of $G_0W_0$ for different exchange-correlation functionals, where for molecular systems Hartree--Fock or functionals with high levels of exact exchange are expected to perform best \cite{Bruneval2013, Jin2019}. This agrees with the finding that $G_0W_0$@HF performs well for the aggregated results across this test set in Fig.~\ref{fig:gw100}. For the IP, we can also compare to self-consistent extensions of $GW$, where they are found to be highly reliable, especially the qs$GW$ method, as found previously \cite{VanSetten2015}. We note that the performance improvements of these self-consistent flavors is not universal (especially in solid-state systems), where a degradation in results has also been found due to favorable cancellation of errors with the neglected vertex term in non-self-consistent versions of $GW$ \cite{Holm1998,Wolf1998,Kutepov2016,Kotliar2017,Berkelbach2019}.}

% ERRORS INCLUDING THE ECP SYSTEMS:
% ---------------------------
% IP    MAE     MSE     STD  
% AGF2  0.501	-0.321	0.816
% ADC2  0.587	0.567	0.897
% ---------------------------
% EA    MAE     MSE     STD  
% AGF2  0.266	-0.158	0.415
% ADC2  0.341	0.234	0.520

The efficient and parallelisable truncation algorithm we choose for each iteration of this AGF2 work proceeds by finding the smallest number of auxiliary states representing the effective self-energy which conserves the zeroth and first spectral moment of the lesser and greater self-energies.
In the first iteration, the auxiliaries representing these self-energies span the $2h1p$ and $1h2p$ spaces respectively, before these are further renormalised to capture implicit higher-order excitations in subsequent iterations.
These spaces are separately rotated and truncated each iteration, before being combined in an effective hermitian Hamiltonian.
A more detailed description of other truncations, including conserving spectral moments in the resulting Greens function\cite{Sriluckshmy2021}, can be found in Ref.~\onlinecite{Backhouse2020a}.
However, in previous work, we showed that the simplest truncation where just the zeroth and first spectral moments of the self-energy are conserved yielded accurate results for the prediction of IPs and EAs of small molecules\cite{Backhouse2020b}, further corroborated by the results of Fig.~\ref{fig:gw100}.
This truncation transforms Eq.~\ref{eq:dyson_eq_eig} into a reduced dimensionality problem,
\begin{align} \label{eq:post_compression_h}
    \begin{bmatrix}
        & F & T^< & T^>& \\
        & T^{<\dagger} & M^< & 0 & \\
        & T^{>\dagger} & 0 & M^>  & 
    \end{bmatrix} \phi = \lambda \phi.
\end{align}
All blocks in the Hamiltonian above have dimensionality of $N$ (compared to $\mathcal{O}[N^3]$ of Eq.~\ref{eq:dyson_eq_eig}), and therefore the resulting Hamiltonian can be completely diagonalized in only $\mathcal{O}[N^3]$ computational effort.

The conserved self-energy spectral moments are defined (for the `lesser' contributions to Eq.~\ref{eq:mp2_self_energy}) as 
\begin{align} \label{eq:mp2_zeroth_moment}
    U_{pq}^{(0)} &= \sum_{ija} (pi|ja) [ 2 (qi|ja) - (qj|ia) ] = v^<v^{<\dagger} \\ \label{eq:mp2_first_moment}
    U_{pq}^{(1)} &= \sum_{ija} (pi|ja) [ 2 (qi|ja) - (qj|ia) ] (\epsilon_i + \epsilon_j - \epsilon_a) \\
    &= v^<E^<v^{<\dagger}.
\end{align}
The required blocks of Eq.~\ref{eq:post_compression_h} can be obtained straightforwardly from these spectral moments, via a single step of a modified block Lanczos approach\cite{Bai1999,Kim1989,Iguchi1992,Weikert1996}.
This can efficiently proceed as a Cholesky decomposition\cite{Fukaya2014,Fukaya2020}, as
\begin{equation}
    \label{eq:block_lanczos_chol_3}
    T^< T^{< \dagger} = U^{(0)},
\end{equation}
with the diagonal blocks subsequently computed as
\begin{equation}
    M^< = (T^{< -1})^{\dagger} U^{(1)} T^{< -1} ,
\end{equation}
with analogous expressions for the auxiliary contributions from the `greater' self-energy.
It should also be noted that the auxiliaries couple to both the particle and hole spaces of the `physical' $F$ block, resulting in mixing between occupied and virtual spaces in the diagonalisation step, despite no direct coupling between the greater and less self-energy contributions.
The highest scaling step of the calculation is the construction of the self-energy spectral moments of Eqs.~\ref{eq:mp2_zeroth_moment} and \ref{eq:mp2_first_moment}. 
With the use of density fitting, this constitutes the only $\mathcal{O}[N^5]$ step each iteration, which can be more accurately divided into a scaling with respect to occupied ($o$) and virtual ($v$) orbitals of $\mathcal{O}[N^2 o^2 v + N^2 o v^2]$, where $N = o + v$.
All other steps, such as the transformation of the density-fitted electronic repulsion integrals or construction of the Fock matrix are lower scaling, at $\mathcal{O}[N^4]$.
%The implementation also maintains an $\mathcal{O}[N^3]$ memory overhead.
%The present implementation also maintains a dependency on memory of $\mathcal{O}[N^3]$, ensuring scalability with respect to available RAM.

\ifjcp
    \begin{table}[htb!]
        \centering
        \includegraphics[width=\columnwidth]{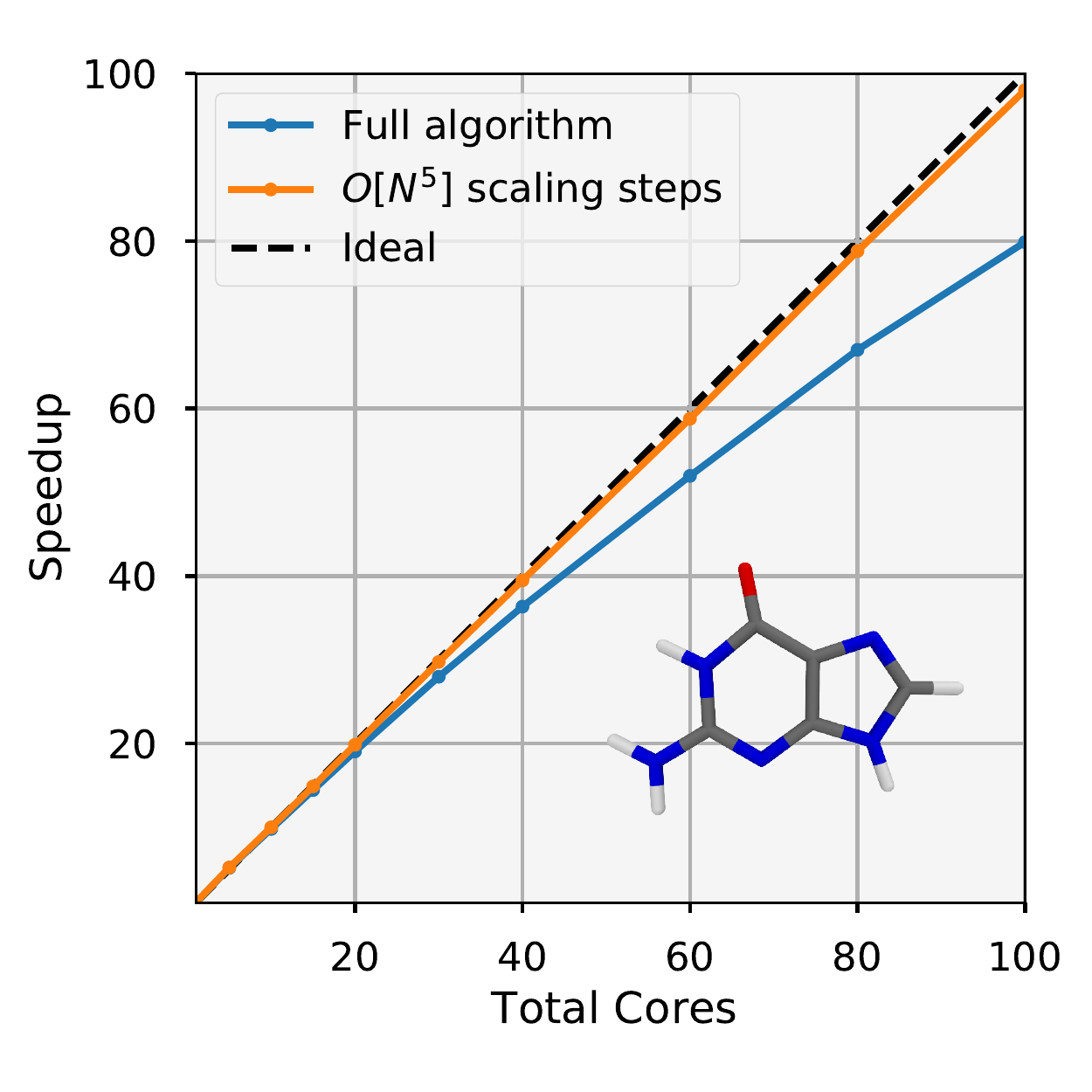}
        \captionof{figure}{Parallel speedup of AGF2 iterations for the guanine molecule using an aug-cc-pVDZ basis with 298 basis functions and 78 electrons correlated.
        The scaling of the $\mathcal{O}[N^5]$ step is shown separately to the full algorithm, where larger systems will be increasingly dominated by this step, indicating an improved parallel scaling as systems become larger.
        The guanine molecule was relaxed using the GFN-xTB method\cite{Grimme2017,Bannwarth2019,Bannwarth2021}, with an aug-cc-pVTZ-RI density fitting basis, with nodes consisting of two 20-core Intel Xeon Gold 6248 2.5 GHz processors.
        On 100 cores, the time per iteration was less than three minutes.
        }
        \label{fig:parallel_speedup}
    \end{table}
\else
    \begin{figure}[tb!]
        \includegraphics[width=0.5\textwidth]{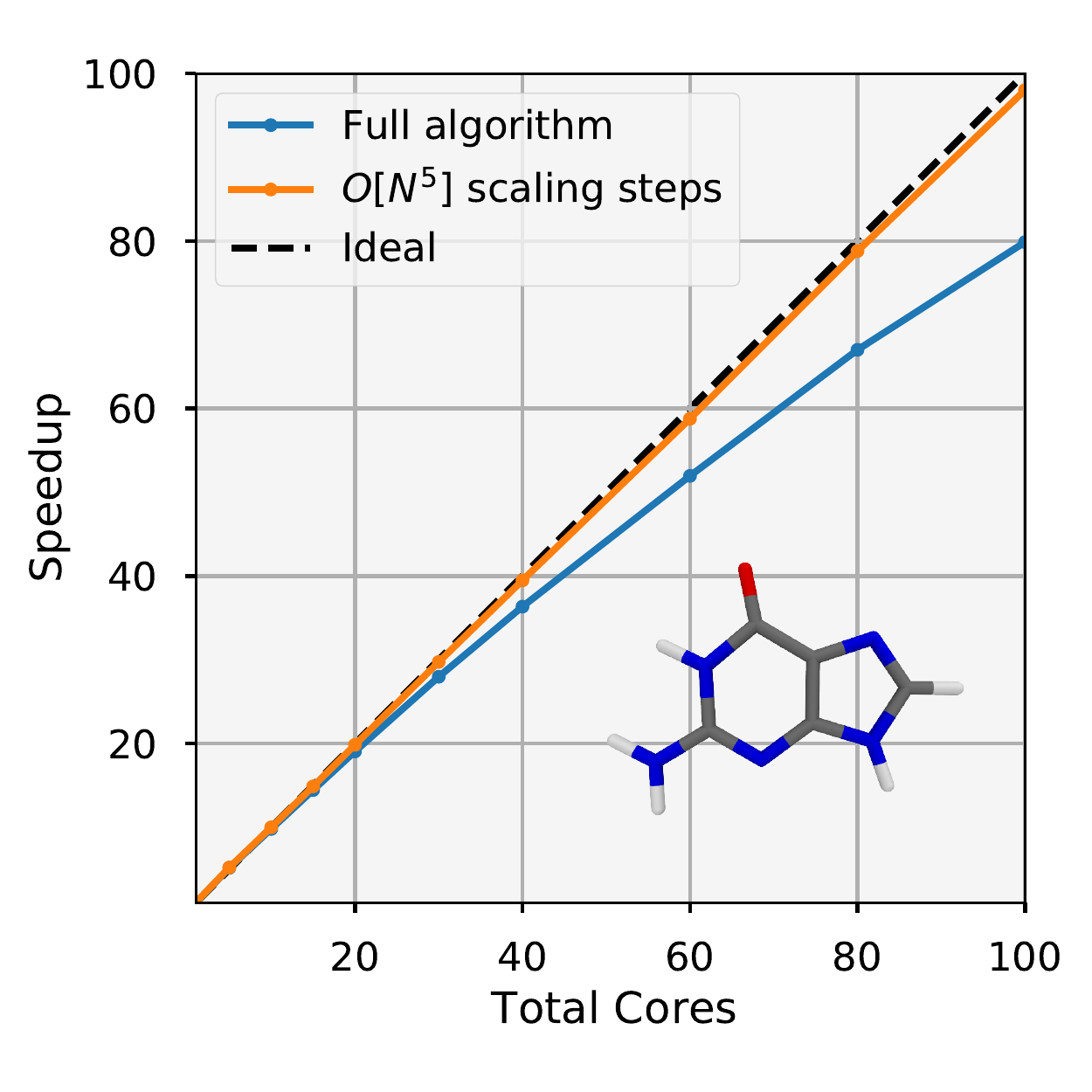}
        \caption{Parallel speedup of AGF2 iterations for the guanine molecule using an aug-cc-pVDZ basis with 298 basis functions and 78 electrons correlated.
        The scaling of the $\mathcal{O}[N^5]$ step is shown separately to the full algorithm, where larger systems will be increasingly dominated by this step, indicating an improved parallel scaling as systems become larger.
        The guanine molecule was relaxed using the GFN-xTB method\cite{Grimme2017,Bannwarth2019,Bannwarth2021}, with an aug-cc-pVTZ-RI density fitting basis, with nodes consisting of two 20-core Intel Xeon Gold 6248 2.5 GHz processors.
        On 100 cores, the time per iteration was less than three minutes.
        }
        \label{fig:parallel_speedup}
    \end{figure}
\fi

\begin{figure}[t!]
    \begin{minipage}{0.625\textwidth}
        \centering
        \includegraphics[width=\textwidth]{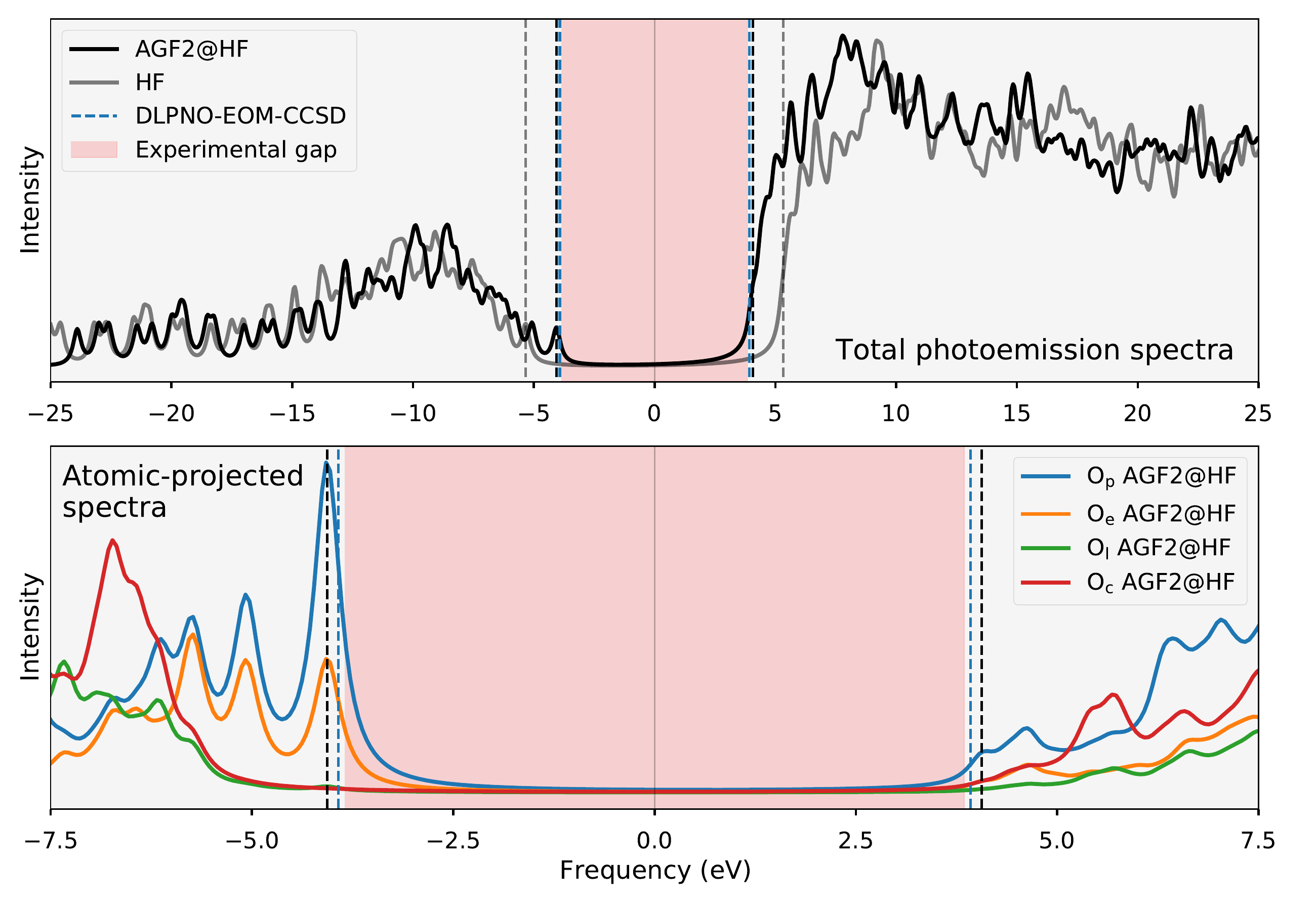}
        \caption{Simulated photoelectron spectra of artemisinin calculated using AGF2 within an aug-cc-pVDZ basis, shifted to line up the centers of the band gaps.
        The geometry was obtained from x-ray crystallographic data from The Cambridge Crystallographic Data Centre, identifier QNGHSU03\cite{Groom2016}.
        The calculation converged in around 16 hours using 8 MPI tasks with 20 OpenMP threads each, for a total of 160 cores.
        The first panel shows the Hartree--Fock spectrum, and that from the subsequent AGF2 calculation.
        The lower panel shows spectral weight projected onto the atomic oxygen atoms of the molecule, over a finer energy resolution, indicating the atomic amplitude of each of the excitations.
        A broadening factor $\eta$ of 0.2 eV was used.
        }
        \label{fig:artemisinin_spectrum}
    \end{minipage}%
    \hspace{\fill}%
    \begin{minipage}{0.35\textwidth}
        \includegraphics[trim={40cm 20cm 40cm 30cm},width=0.8\textwidth,clip]{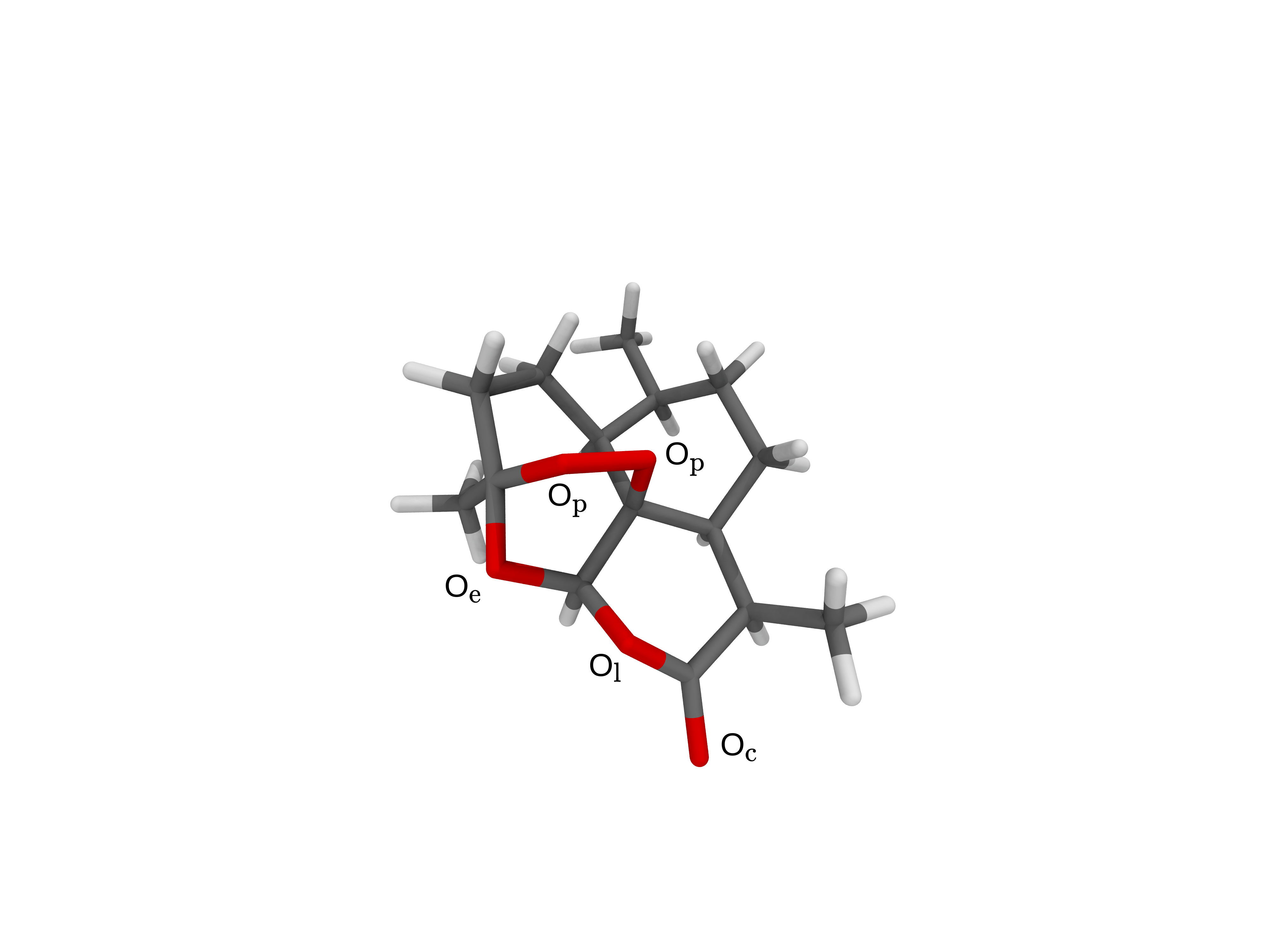}
        \caption{Artemisinin molecular structure, with the different oxygen types labelled as O$_\mathrm{p}$, O$_\mathrm{e}$, O$_\mathrm{l}$ and O$_\mathrm{c}$, denoting the peroxidic, etheric, inner-ring lactonic and lactonic carbonyl atoms respectively.}
        \label{fig:artemisinin_labelled}
        \includegraphics[trim={40cm 20cm 40cm 30cm},width=0.8\textwidth,clip]{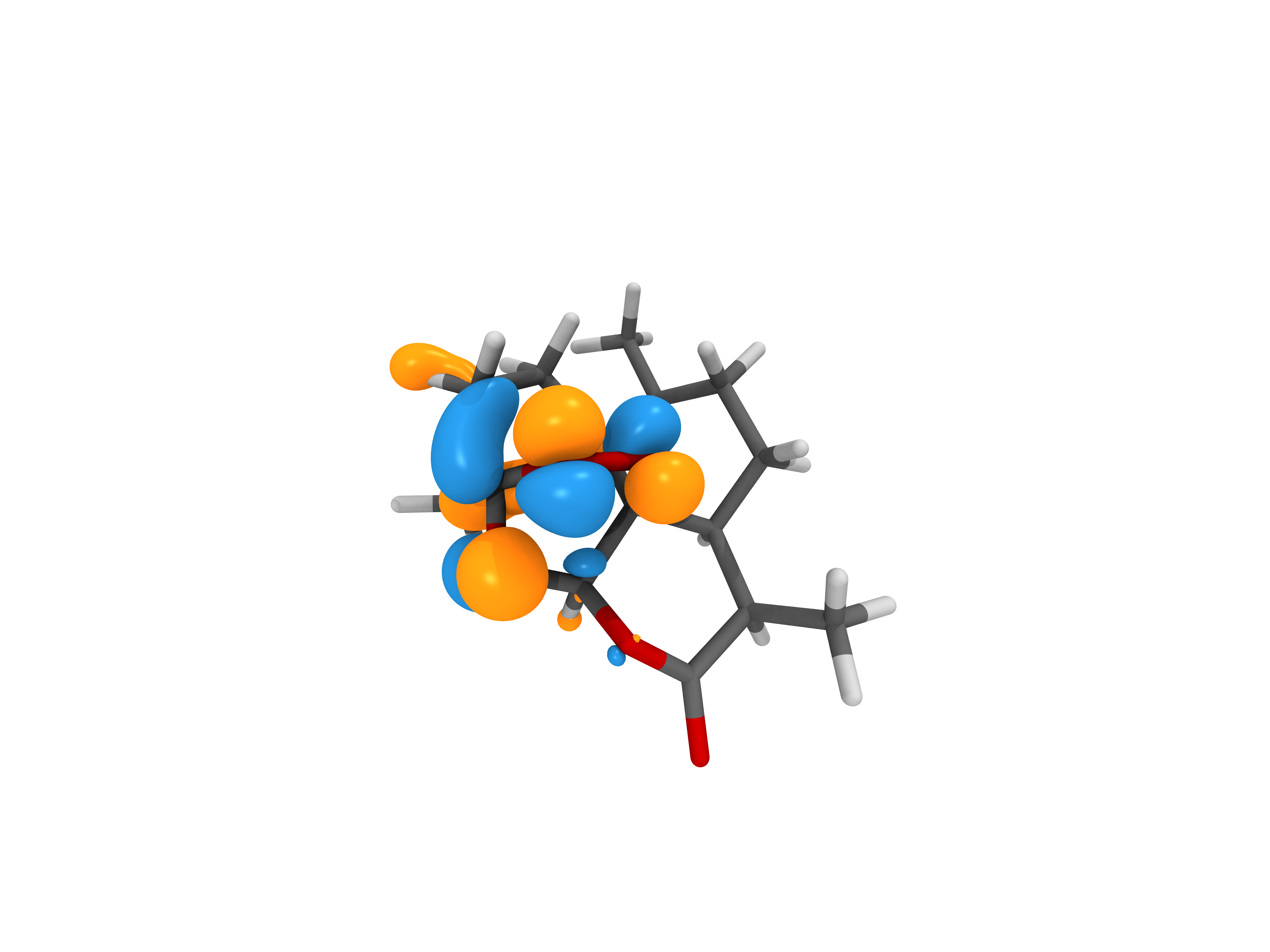}
        \caption{Spatial distribution of the Dyson orbital corresponding to the first ionization potential of the artemisinin molecule, plotted with an isovalue of 0.05.
        }
        \label{fig:artemisinin_hoqmo}
    \end{minipage}
\end{figure}

Importantly, the computational bottleneck represented by Eqs.~\ref{eq:mp2_zeroth_moment} and \ref{eq:mp2_first_moment} can be very efficiently parallelised by distributing blocks of the summed indices onto MPI processes, with OpenMP parallelism within each blocks introducing a hybrid parallel algorithm for high-performance computing environments over distributed memory.
This scheme has a very low communication overhead, with each MPI thread exploiting independent matrix multiplications via optimised libraries\cite{Xianyi2012}.
Fig.~\ref{fig:parallel_speedup} shows the parallel scaling of the resulting algorithm for a guanine DNA nucleobase in an aug-cc-pVDZ basis, demonstrating a perfect parallelism for the dominant $\mathcal{O}[N^5]$ scaling steps.
At larger core counts, the less effective parallelism of the density-fitting integral transformations and Fock matrix construction become noticeable, however for larger systems these are expected to become an increasingly small fraction of the total resources required.
Further details on this efficient hybrid parallel scheme are included in the supplementary information.
This is implemented in the open-source {\tt PySCF} simulation package\cite{pyscf}, where the AGF2 code can scale to beyond 1000 orbitals (requiring approximately 5 days on 160 cores), with input files used for the generation of these results also included in the supplementary information. 
%This is particularly convenient on high-performance computing clusters, in which MPI processes may be distributed between nodes, and OpenMP threads spawned within each node.
%The computations within each block can be written largely as calls to \texttt{DGEMM}, meaning the majority of operations are spent inside highly optimized libraries such as \texttt{LAPACK} and \texttt{OpenBLAS}\cite{TODO}.
%This layout combines reasonably efficient parallelism with ease of implementation and code maintenance.

%Fig.~\ref{fig:parallel_speedup} shows the efficacy of this layout for the first three iterations of a guanine molecule in an aug-cc-pVDZ basis.
%Since the construction of Eqs.~\ref{eq:mp2_zeroth_moment} and \ref{eq:mp2_first_moment} are the only $O(N^5)$ sections of the density-fitted calculation, their parallelization is paramount for the scalability of the code.
%\note{Include a little picture of the guanine system in the plot}
%\note{Do we want to also show a hybrid calculation?}
%\note{Should I remove the $O(N^5)$ curve?}
%\note{There is some superlinearity - is this easily explainable? It may be inconsistencies in performance, these are single calculations so maybe I ought to run a few and take an average?}

%
% ARTEMISININ
%

% Novak and Kovac: https://pubs.acs.org/doi/pdf/10.1021/jo030077g
% Galasso: https://doi.org/10.1016/j.chemphys.2007.04.008

Artemisinin is a molecule of particular interest due to its antimalarial activity, with the discovery of artemisinin-combination therapies in 1972 earning co-receipt of the Nobel Prize in Medicine in 2015\cite{Qinghaosu1979,White2015}.
The site and energy of electron addition and removal in this C$_{15}$H$_{22}$O$_{5}$ molecule is critical in a description of its resulting activity, where a reactive endoperoxide bond leaves the molecule susceptible to electrophilic attack. It is thought that the dissociative reduction of this bond and subsequent radicals generated by this process are key to the antimalarial properties of the drug which ultimately destroy the parasite.
Using UV photoelectron spectroscopy (UV-PES), Novak and Kova{\u c} identified the ionization potential of artemisinin to be 9.40~eV\cite{Novak2003}.
%, with an additional neighbouring peak at 9.75 eV\cite{Novak2003}.
Density functional theory (DFT) calculations with the B3LYP functional were used in order to support the assignment of these bands via Koopmans theorem, along with further experimental UV-PES on different fragments of the molecule. These attributed the spatial character of the ionization to the oxygen lone pairs of this endoperoxide bond, i.e. the ionization having $n(\mathrm{O_p})^{-}$ character, consistent with the main reactive site of the molecule.
%rather than lone pairs on the carbonyl oxygen of the lactone ring.
However in later photoelectron experiments by Galasso \textit{et al.}, they corroborated the ionization potential of Novak and Kova{\u c} (revising it to 9.75~eV) but differed in their assignment of its spatial location\cite{Galasso2007}.
%While assigning a value of 9.45 eV for the IP, they also identified an unresolved signal at 9.8 eV in agreement with the second peak in the analysis by Novak and Kova{\u c}.
Supplementing their experiment with numerical OVGF calculations, they found the ionization to be associated with two near-degenerate states with $n(\mathrm{O_c,O_l})$ and $n(\mathrm{O_p,O_c,O_e,O_l})$ character respectively, where $\mathrm{O_p,O_c,O_e,O_l}$ indicate the peroxidic, lactonic carbonyl, etheric and inner-ring lactonic oxygen sites, respectively, as shown in Fig.~\ref{fig:artemisinin_labelled}.
This more delocalized and degenerate state differed from the previous assignment of the primary ionization site of the molecule.
Galasso \textit{et al.} additionally probed the electron affinity using electron transmission spectroscopy, which they found to have an energy of 1.76 eV.
They again supplemented this finding with OVGF calculations and assigned the excitation to primarily $\sigma^*(\mathrm{O_pO_p})$ orbital character, with additional contributions from $\pi^*(\mathrm{CO_c})$.
These observations provided an experimental fundamental bandgap of 7.69~eV, while their supporting OVGF calculations gave a substantially underestimated bandgap of 5.4~eV \cite{Galasso2007}.
% HF gap: 10.67 eV

We compute the single-particle spectrum at the AGF2 level for this system, in a aug-cc-pVDZ basis of 658 orbitals and 152 correlated electrons, requiring 2,500 CPU hours to fully converge.
In Fig.~\ref{fig:artemisinin_spectrum} we show excellent agreement with the experimental band gap, obtaining a gap of 8.13~eV and successfully relaxing the inaccurate Hartree--Fock reference spectrum. Comparison spectra from ground-state density functional theory at the level of PBE gives 2.11~eV and B3LYP gives 4.36~eV, significantly underestimating the gap\cite{Cohen2012}.
\toadd{
    An analysis at the DLPNO-EOM-CCSD level of theory provides a value for the gap of 7.85~eV, calculated using the {\tt ORCA} program package\cite{Neese2012,Neese2020,Riplinger2013,Riplinger2016}, which is expected to be a good approximation to the full EOM-CCSD results. This close agreement of the gap with the AGF2 value (0.28~eV discrepancy), provides evidence that the accuracy from the GW100 test set transfers to larger applications. 
}
By projecting the converged highest occupied quasi-molecular orbital into the physical space, we can resolve the spatial Dyson orbital corresponding to the first ionization, shown in Fig.~\ref{fig:artemisinin_hoqmo}.
This peak is relatively isolated in the spectrum as a single state, with a quasiparticle weight of 0.947.
The dominant contributions to this excitation arise from both the peroxidic and etheric oxygen lone pairs, with smaller additional contributions from the carbon atoms neighbouring them.
This is in agreement with the assignment by Novak and Kova{\u c}, while the lactone group highlighted as the primary ionization site in the work of Galasso \text{et al.} has insignificant contribution.
%The ability to assign non-unit transition amplitudes to these peaks highlights an advantage of the AGF2 method over mean-field approaches, as the auxiliary states remove spectral weight from the physical space.
%Galasso \text{et al.} further assign lone pair contributions from the etheric and peroxidic oxygens, however, which is observed in the AGF2 results.
%The additional unresolved experimental peak of Galasso \text{et al.} assigned to $\sigma(\mathrm{CCO_p})$ however agrees with these AGF2 results, which corresponds to the second ionization potential shown in Fig.~\ref{fig:artemisinin_spectrum}, found .

The electron-attachment states of AGF2 form a dense manifold of overlapping and relatively low-weight transitions, such that describing the character of these states via a projected spectrum onto the different atomic contributions provides a better reflection of the character of these excitations than any single state-specific description. This is shown in the lower plot of Fig.~\ref{fig:artemisinin_spectrum}, projecting onto atomic local meta-Lowdin states\cite{Sun2014}, indicating that the dominant oxygen contributions from the lowest unoccupied states is characterized by the endoperoxidic antibonding orbital, in agreement with Galasso \textit{et al.}, however significant low-energy contributions also arise from the ether group. The specific Dyson orbital corresponding to the electron attachment state is shown in the supplementary information. Overall, the character of these states supports the view that initial reduction of the artemisinin proceeds via the endoperoxidic oxygen sites.

In conclusion, in this work we have presented an approach to the computation of the full charged excitation spectrum of correlated molecules. This relies on a combination of many-body perturbation theory complete through second order, with self-consistent iteration of these diagrams to dress the electron propagators to infinite order. As a way to improve both the numerical accuracy of the resulting spectra, as well as reducing computational effort, we additionally perform a non-perturbative truncation of the resulting effective self-energy at each renormalization step of the propagators, while conserving the first two spectral moments of the self-energy. This results in an efficient algorithm which nevertheless performs very favourably when compared to similarly scaling methods, and admits a large-scale parallel algorithm, now publicably available within the {\tt PySCF} simulation package. Application and interpretation of biologically relevant molecules such as the artemisinin study in this work, pave the way for further investigations into this emerging approach for charged spectra.

\ifjcp\else
    \newpage
\fi

\section*{Acknowledgements}

The authors sincerely thank Xavier Blase and Ivan Duchemin for help obtaining $G_{0}W_{0}$ results over the GW100 test set, as well as Alexander Sokolov for advice on ADC and helpful suggestions on the manuscript.
G.H.B. also gratefully acknowledges support from the Royal Society via a University Research Fellowship, as well as funding from the European Research Council (ERC) under the European Union’s Horizon 2020 research and innovation programme (Grant Agreement No. 759063).
We are grateful to the UK Materials and Molecular Modelling Hub for computational resources, which is partially funded by EPSRC (EP/P020194/1).

\toadd{
    \section*{Supporting Information}
    
    Supporting information includes the IP and EA values at the AGF2/def2-TZVPP level for the $GW100$ database, as well as the electron-attached Dyson orbital for the Artemisinin, coordinates for this molecule, and an example {\tt PySCF} input file for reproducing all AGF2 results. Further details on the parallelization strategy are also given.
}

\ifjcp
%    \bibliographystyle{acs}
    %\bibliography{references}
\else
    %\section{References}
    %\bibliographystyle{achemso}
    %\bibliography{references}
\providecommand{\latin}[1]{#1}
\makeatletter
\providecommand{\doi}
  {\begingroup\let\do\@makeother\dospecials
  \catcode`\{=1 \catcode`\}=2 \doi@aux}
\providecommand{\doi@aux}[1]{\endgroup\texttt{#1}}
\makeatother
\providecommand*\mcitethebibliography{\thebibliography}
\csname @ifundefined\endcsname{endmcitethebibliography}
  {\let\endmcitethebibliography\endthebibliography}{}

\fi

\end{document}